\documentclass{article}

\usepackage[english]{babel}

\usepackage[a4paper,top=2cm,bottom=2cm,left=3cm,right=3cm,marginparwidth=1.75cm]{geometry}

\usepackage{amsmath,amssymb}
\usepackage{siunitx}
\PassOptionsToPackage{hyphens}{url}\usepackage{hyperref}
\usepackage{cleveref}
\usepackage[utf8]{inputenc}
\usepackage{csquotes}
\usepackage{booktabs}
\usepackage{longtable}
\usepackage{adjustbox}
\usepackage{array}
\usepackage{url}
\usepackage{titlesec}
\usepackage{authblk}
\usepackage{xcolor} % Load the xcolor package for color options

\titleformat{\subsection}
  {\mdseries\itshape\large} % Medium series, italic shape, and large font size
  {\thesubsection}{1em}{} % Numbering, spacing, and the section title itself

\usepackage[english]{babel}
\usepackage[style=authoryear,backend=bibtex,natbib=true,maxcitenames=2,uniquelist=false]{biblatex}
\DeclareNameAlias{sortname}{family-given}
\DeclareNameAlias{default}{family-given}

\renewbibmacro{in:}{}
\DeclareFieldFormat[article]{title}{\mkbibquote{#1}\addcomma}
\DeclareFieldFormat[book]{title}{\mkbibemph{#1}\addcomma}
\DeclareFieldFormat[bookinbook]{title}{\mkbibemph{#1}\addcomma}
\DeclareFieldFormat[inbook]{title}{\mkbibquote{#1}\addcomma}
\DeclareFieldFormat[incollection]{title}{\mkbibquote{#1}\addcomma}
\DeclareFieldFormat[inproceedings]{title}{\mkbibquote{#1}\addcomma}
\DeclareFieldFormat[manual]{title}{\mkbibemph{#1}\addcomma}
\DeclareFieldFormat[misc]{title}{\mkbibemph{#1}\addcomma}
\DeclareFieldFormat[thesis]{title}{\mkbibemph{#1}\addcomma}
\DeclareFieldFormat[unpublished]{title}{\mkbibquote{#1}\addcomma}
\DeclareFieldFormat[patent]{title}{\mkbibemph{#1}\addcomma}
\DeclareFieldFormat[report]{title}{\mkbibemph{#1}\addcomma}
\DeclareFieldFormat[online]{title}{\mkbibquote{#1}\addcomma}
\DeclareFieldFormat[software]{title}{\mkbibemph{#1}\addcomma}
\DeclareFieldFormat[booklet]{title}{\mkbibemph{#1}\addcomma}
\DeclareFieldFormat[periodical]{title}{\mkbibemph{#1}\addcomma}
\DeclareFieldFormat[standard]{title}{\mkbibemph{#1}\addcomma}

\DeclareFieldFormat[article]{journaltitle}{\iffieldundef{shortjournal}{\mkbibemph{#1}\addcomma}{\mkbibemph{\printfield{shortjournal}}\addcomma}}
\DeclareFieldFormat{volume}{\bibstring{volume}~#1}
\DeclareFieldFormat{number}{\bibstring{number}~#1}

\DefineBibliographyStrings{english}{
  volume = {Vol.},
  number = {No.}
}

\renewbibmacro*{volume+number+eid}{%
  \printfield{volume}%
  \setunit*{\addspace}%
  \printfield{number}%
  \setunit{\addcomma\space}%
  \printfield{eid}}

\renewbibmacro*{journal+issuetitle}{%
  \usebibmacro{journal}%
  \setunit*{\addcomma\space}%
  \usebibmacro{volume+number+eid}%
  \setunit{\addcomma\space}%
  \usebibmacro{issue+date}}

\renewbibmacro*{publisher+location+date}{%
  \printlist{publisher}%
  \iflistundef{location}
    {\setunit*{\addcomma\space}}
    {\setunit*{\addcolon\space}}%
  \printlist{location}%
  \setunit*{\addcomma\space}%
  \usebibmacro{date}}

\DeclareCiteCommand{\cite}[\mkbibparens]
  {\usebibmacro{prenote}}
  {\usebibmacro{citeindex}%
   \usebibmacro{cite}}
  {\multicitedelim}
  {\usebibmacro{postnote}}

\renewbibmacro*{cite:labelyear+extrayear}{%
  \iffieldundef{labelyear}
    {}
    {\printtext[bibhyperref]{%
       \printfield{labelyear}%
       \printfield{extrayear}}}}

\renewbibmacro*{cite:labeldate+extradate}{%
  \iffieldundef{labelyear}
    {}
    {\printtext[bibhyperref]{%
       \printfield{labelyear}%
       \printfield{extradate}}}}

\AtEveryBibitem{
  \clearfield{month}
  \clearfield{day}
  \ifentrytype{book}{
    \clearlist{location}
  }{}
}

\DefineBibliographyStrings{english}{
  andothers = {\textit{et al.},}
}

\DeclareFieldFormat[article]{volume}{\bibstring{jourvol}\addnbspace #1}
\DeclareFieldFormat[article]{number}{\bibstring{number}\addnbspace #1}
\DeclareFieldFormat[article]{volume}{Vol. #1}
\DeclareFieldFormat[article]{number}{No. #1}
\DeclareFieldFormat{url}{\bibstring{available at}\addcolon\space\url{#1}}
\DeclareFieldFormat{urldate}{\mkbibparens{accessed \addspace#1}}

\DeclareFieldFormat{urldate}{%
  \mkbibparens{accessed\space%
    \thefield{urlday}\addspace%
    \mkbibmonth{\thefield{urlmonth}}\addspace%
    \thefield{urlyear}}}

\crefformat{figure}{#2Figure~#1#3}
\Crefformat{figure}{#2Figure~#1#3}
\crefformat{table}{#2Table~#1#3}
\Crefformat{table}{#2Table~#1#3}
\crefformat{section}{#2Section~#1#3}
\Crefformat{section}{#2Section~#1#3}

\author[1]{Shuai Huang}
\author[1]{Thomas Denney}
\author[2]{Deqiang Qiu\thanks{ Corresponding author: Deqiang Qiu (deqiang.qiu@emory.edu)}}

\affil[1]{Auburn University, Auburn, US}
\affil[2]{Emory University, Atlanta, US}

\title{Improving the Test-retest Reliability of Quantitative Susceptibility Mapping}

\begin{document}
\maketitle

\begin{abstract}
\textbf{Motivation} - The test-retest reliability of quantitative susceptibility mapping (QSM) is affected by parameters of the acquisition protocol such as the angulation of acquisition plane with respect to the B0 field direction and spatial resolution.

\textbf{Goal} - We aim to reduce the protocol-dependent biases/errors that might overshadow subtle changes of the susceptibility values due to pathology in the brain.

\textbf{Approach} - The magnitude and phase images used for QSM are registered according to a standard reference protocol in the k-space through nonuniform fast Fourier transform (NUFFT).

\textbf{Results} - With the help of our proposed approach, the test-retest reliability of scans acquired from different protocols is notably improved.

\textbf{Impact} - The improved test-retest reliability of QSM makes it easier to detect subtle susceptibility changes in the early stages of neurodegeneration such as Alzheimer's disease and Parkinson's disease.
\\
\\
%add 6 keywords
%\textbf{Keywords:} keyword 1; keyword 2; keyword 3, keyword 4; keyword 5; keyword 6
\end{abstract}
%\linenumbers

%\section*{Template Overview}
%\textcolor{blue}{This \LaTeX template is designed to incorporate the specific Harvard citation style defined by Emerald Publishing, along with section styles and other adjustments required for submission in Construction Innovation journal.
%Users should adapt the styling to align with the guidelines of the journal to which they are submitting their article. Make sure to remove the authors' names and acknowledgements if you are submitting an anonymous file for double-blind peer review.}

\section{Introduction}
\label{sec:introduction}

QSM recovers the susceptibility map of tissue from a phase image. It is widely used to study iron deposition or pathologies such as hemorrhage and calcification in the brain. The acquisition protocol mainly affects the reconstructed susceptibility values through two parameters: the angulation of the acquisition plane with respect to the main magnetic field $B_0$ and the spatial resolution of the voxel. In clinical scans, oblique acquisitions are commonly used to increase the anatomical coverage and to align the image axes with the anterior commissure-posterior commissure axis. Additionally, anisotropic image resolution are commonly used because of scan time and SNR considerations. These practices cause inaccuracy in QSM quantification as detailed below. First, we solve an ill-posed dipole inversion problem to recover the susceptibility map from the local field map. The dipole kernel is a function of the $B_0$ field direction $\overrightarrow{\boldsymbol b_0}$ in the coordinate system defined by the image axes rather than the scanner axes. As a result, the dipole kernel of an oblique acquisition is different from that of a non-oblique (straight) acquisition. This creates an inherent bias in the QSMs from oblique acquisitions. Second, the values of raw phase image are within $[-\pi,\pi)$ and need to be unwrapped to produce a continuous phase image. The unwrapped results from raw phase images with isotropic spatial resolution are typically more accurate than those with anisotropic resolution. To improve the test-retest reliability of QSM, we propose an approach that minimizes the biases/errors introduced by the acquisition protocol.

\section{Method}
\label{sec:method}
The protocol-dependent biases/errors are minimized by registering the MR images into a common space defined by a reference protocol. In particular, the reference protocol is chosen to produce a non-oblique acquisition with isotropic spatial resolutions. Kiersnowski et al. registered the unwrapped phase images directly in the image space \cite{Kiersnowski:2023}. However, this method introduces additional interpolation errors that are passed on to the susceptibility maps, and such errors get larger for phase images with anisotropic resolutions. 

\begin{figure}[tpb!]
\begin{center}
\includegraphics[width=\textwidth]{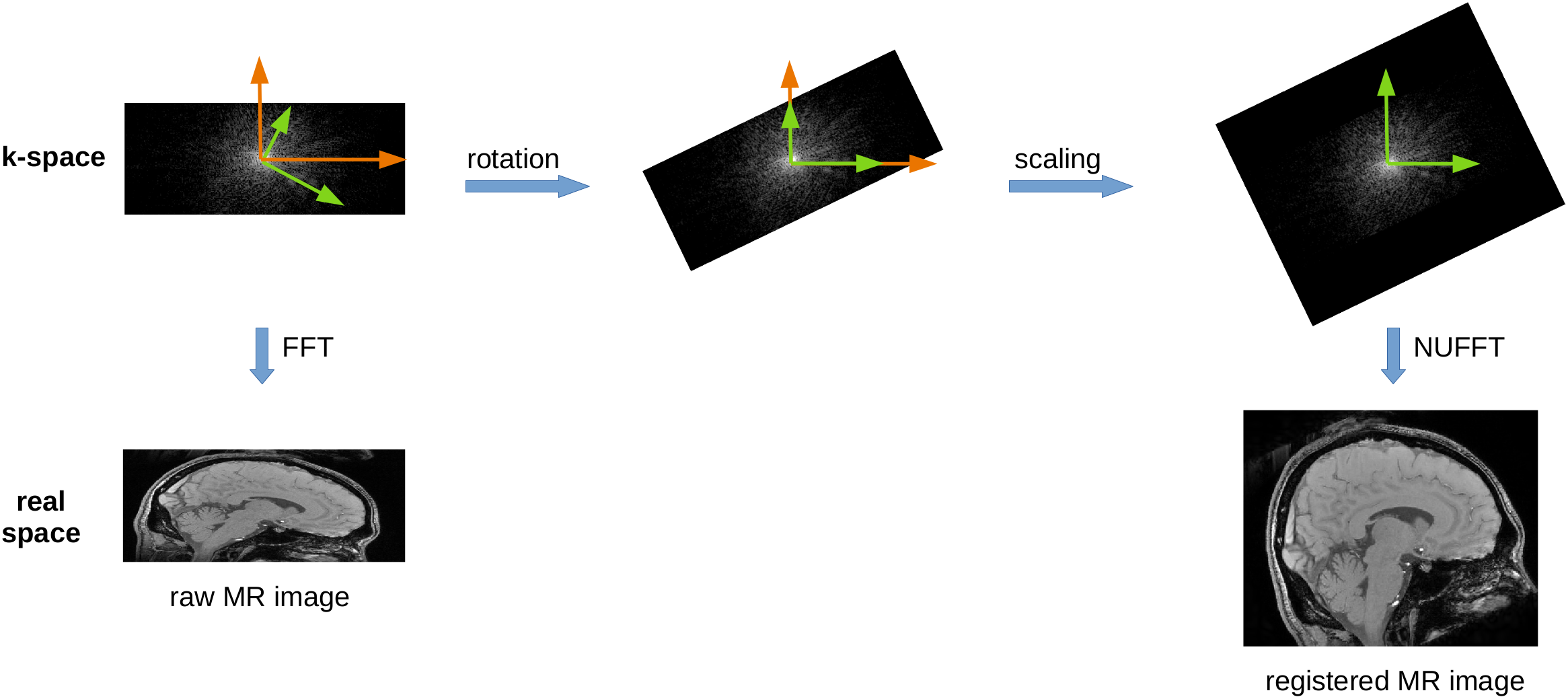}
\caption{$k$-space registration according to a reference protocol with non-oblique acquisition and isotropic spatial resolutions. The k-space data is rotated so that the transformed scanner axes in the $k$-space is aligned with the $k$-space axes. The $k$-space axes are then scaled to produce reconstructions with isotropic spatial resolutions. NUFFT is finally used to reconstruct the registered MR image.}
\label{fig:k_space_registration}
\end{center}
\end{figure}

We propose to perform the registration in the $k$-space in two steps as shown in Fig. \ref{fig:k_space_registration}:
\begin{enumerate}
    \item Rotate the transformed scanner axes in the $k$-space to align them with the $k$-space axes. Let $\boldsymbol l = [l_x\ l_y\ l_z]^T$ denote the location of a Fourier measurement, $\boldsymbol R$ the rotation matrix. The measurement location after the rotation is $\boldsymbol l_R=\boldsymbol R\boldsymbol l$.
    \item Scale the $k$-space axes to produce reconstructions with isotropic spatial resolutions. Let $v_x, v_y, v_z$ denote the spatial resolutions of the MR image, $v_r$ the isotropic resolution in the reference protocol. The scaling factors $\boldsymbol s$ on the measurement location are: $\boldsymbol s=[v_r/v_x\ v_r/v_y\ v_r/v_z]^T$. The measurement location after the scaling is $\boldsymbol l_S = \textnormal{diag}\{\boldsymbol s\}\boldsymbol l_R$.
\end{enumerate}
Nonuniform fast Fourier transform (NUFFT) \cite{KeKuPo09} is used to reconstruct the registered MR image from the rotated and scaled $k$-space data. Compared to the previous image space registration approach, the proposed $k$-space registration approach does not involve interpolation and can be applied on generic images with anisotropic resolutions. 

\section{Results}
\label{sec:results}
\begin{figure}[tpb!]
\begin{center}
\includegraphics[width=\textwidth]{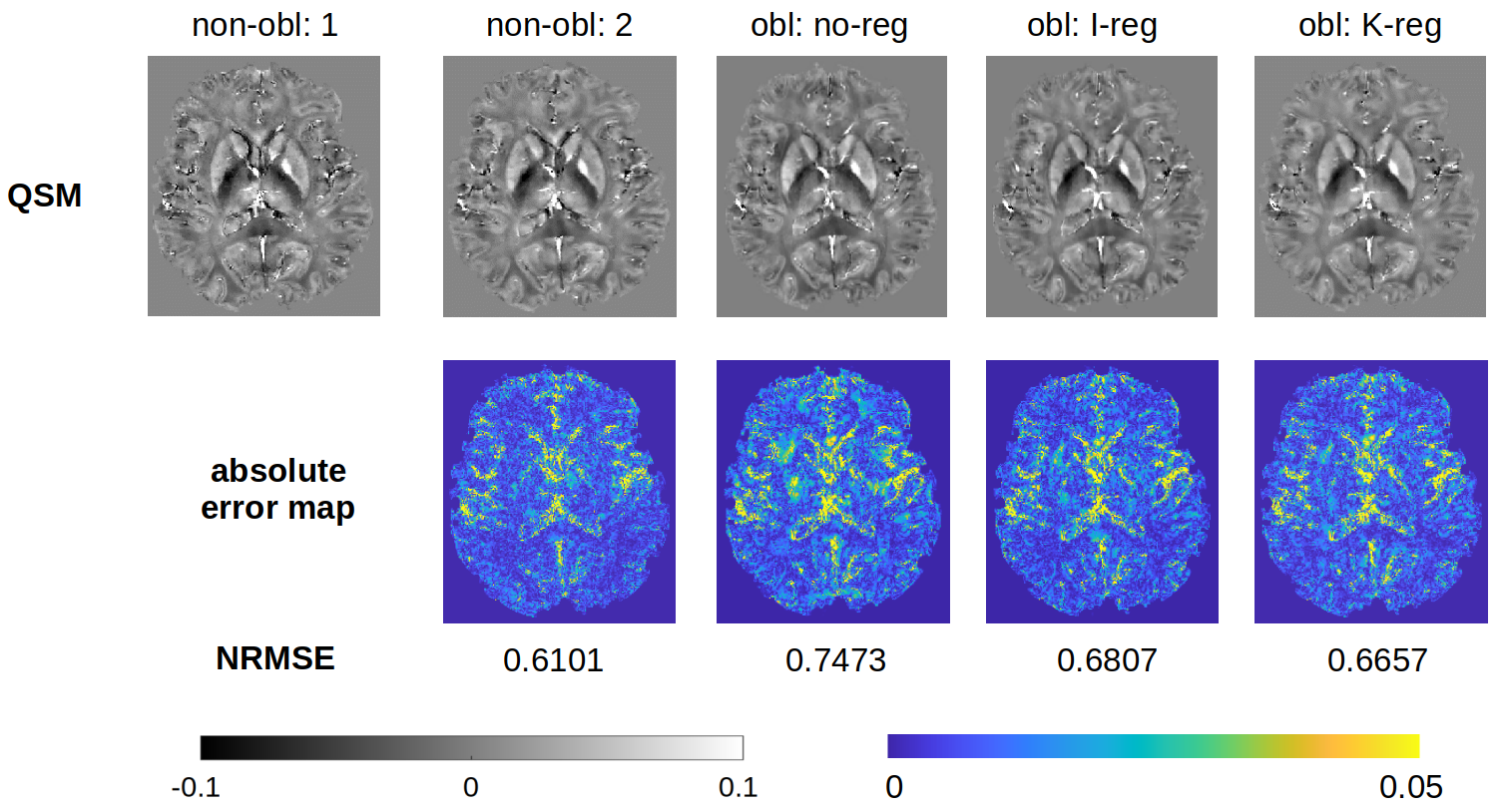}
\caption{The recovered QSMs and the corresponding error maps. The $k$-space measurements are acquired with 3 protocols. The first and second protocols performed the same non-oblique acquisitions with isotropic resolutions, corresponding to the above maps in the first and second columns, i.e. ``non-obl:1'' and ``non-obl:2''. The QSM from the first protocol was used as the reference to calculate the absolute error maps. The third protocol performed oblique acquisition with anisotropic resolution: the third column shows the QSM recovered without phase-image registration (``obl: no-reg''); the fourth column shows the QSM recovered with the image-space registration (``obl: I-reg''); the fifth column shows the QSM recovered with the $k$-space registration (``obl: K-reg'').}
\label{fig:rec_qsm}
\end{center}
\end{figure}

The 3D $k$-space measurements were acquired using three different protocols on a 3T MRI scanner (Siemens Prisma). The first and second protocols were the same with non-oblique acquisitions and $0.7$mm isotropic resolutions. The QSM from the first protocol is used as the reference. The differences between the QSMs from the first and second protocols did not contain protocol-dependent biases/errors, and were used to establish the test-retest reliability baseline. The third protocol produced an oblique acquisition with $0.7\times 0.7$mm$^2$ in-plane resolutions and 1.4mm slice thickness. Using the data from the third protocol, we reconstructed three QSMs from the unregistered phase images, the image-space-registered (unwrapped) phase images and the $k$-space-registered phase images respectively. The differences between the QSMs from the first and third protocols were used to evaluate the test-retest reliabilities of the image-space and $k$-space registration methods. The rest parameters were shared across the three protocols: 3 echoes with the first TE=7.94ms and echo spacing=8ms; TR=29ms; pixel bandwidth=260Hz, FoV=224mm. The QSMs were reconstructed using the MEDI toolbox \cite{NonlinearMEDI:Liu:2013} and shown in Fig. \ref{fig:rec_qsm}.

\section{Discussion}
\label{sec:discussion}
The second protocol with non-oblique acquisition produced the lowest NRMSE, which serves as the baseline of test-retest reliability. When it comes to the third protocol with oblique acquisition, both the image-space registration (I-reg) and $k$-space registration (K-reg) approaches produced lower NRMSE than the approach without phase-image registration (no-reg). This is caused by the bias in the oblique dipole kernel used by the no-reg approach. Additionally, due to the interpolation error introduced by the I-reg approach, the recovered QSM from I-reg had higher NRMSE than that from K-reg. The proposed K-reg approach is able to reduce the NRMSE further than the I-reg approach. By registering the raw MR images into a common space, we can improve the test-retest reliability of QSMs from different protocols. This makes it easier to detect subtle susceptibility changes in the early stages of neurodegeneration, which could lead to new research findings that were previously overshadowed by protocol-dependent biases/errors.

%\section{Conclusion}
%\label{sec:conclusion}

\hfill

%\textit{Acknowledgments} \\Mention all external funding sources in the acknowledgements.\\

%\textit{Disclosure statement:} \\No potential conflict of interest is reported by the authors

%\break
\printbibliography

\end{document}